\documentstyle[12pt,psfig]{article}
\def\reference{\parskip 0pt\par\noindent\hangindent 0.5 truecm}

\newcommand{\kms}{km\,s$^{-1}$}

\newcommand{\Msun}{M$_{\odot}$}

\newcommand{\HII}{$\mathrm{H\,{\scriptstyle II}}$}
\newcommand{\NII}{$\mathrm{N\,{\scriptstyle II}}$}

\def\arcmin{\hbox{$^\prime$}}
\def\arcsec{\hbox{$^{\prime\prime}$}}

\newcommand{\CO}{$^{12}${\rmfamily CO}{(1--0)}}

\begin{document}

\title{An investigation of the molecular clouds of the Carina
\HII\ region/molecular cloud complex - First results}

\author{K. J. Brooks $^{1}$ \and
 J. B. Whiteoak $^{2}$ \and
 J. W. V. Storey $^{1}$ 
} 

\date{}
\maketitle

{\center
$^1$ School of Physics, The University of NSW, Sydney, NSW, 2052, Australia\\kbrooks@wodin.phys.unsw.edu.au\\[3mm]
$^2$ Australia Telescope National Facility, CSIRO, PO Box 76, Epping, NSW,
2121, Australia\\[3mm]
}

\begin{abstract}
The Carina Nebula is an extremely bright southern \HII\ region
embedded in a giant molecular cloud and contains some of the most
massive stars known in our Galaxy. We are undertaking a
multi-wavelength study of the Carina Nebula in order to examine the
detailed kinematics and distribution of the molecular and ionised gas,
and to look for further evidence of ongoing star formation. Here we
present the results of the initial molecular cloud observations which
were made by observing the \CO\ emission with the Mopra
antenna. The observations reveal the clumpy morphology of the
molecular gas, and allow us to identify many interesting regions for
follow-up observations.

\end{abstract}

{\bf Keywords:}

\HII\ regions --- ISM: clouds, kinematics and dynamics --- stars: formation

\bigskip

\section{Introduction}

The Carina \HII\ region/molecular cloud complex is an excellent region for
studying the interaction of massive stars with their parental Giant
Molecular Cloud (GMC). The nebula covers an area of $\approx$ 4 deg$^{2}$
and is bisected by a prominent V-shaped dark lane. There are over 14 star
clusters in this region which have been studied extensively over the
past twenty years. For excellent reviews see Feinstein (1995) and Walborn
(1995). The most influential clusters of the nebula are the two OB
clusters, Tr 14 and Tr 16. These clusters contain numerous O-type stars,
including three O3 stars each, making them two of the most massive star
clusters in our galaxy. Tr 14 is a compact cluster situated to the
north-west of the nebula, adjacent to the western dust lane. Tr 16 is an
open cluster centred northwards of the vertex of the dark lane. It contains
one of the most massive stars known: $\eta$ Car. Here we will adopt the
popular view (e.g. Tovmassian 1995, Walborn 1995) that Tr 14 and Tr 16 are
at a common distance of about 2.2 kpc and that Tr 14 is younger than Tr 16.

Considering the extensive studies on the stellar content of the Carina
Nebula, in particular $\eta$ Car and its surrounding Homunculus nebula,
relatively little work has been done on the extended nebula in the last
fifteen years. Early radio continuum observations revealed that the nebula
contains a large ionised region with two peaks, Car I and Car II (Gardner
\& Morimoto 1968). Higher resolution radio continuum data show that both Car
I and Car II are made up of a number of filamentary arcs and rings and are
everywhere thermal (Retallack 1983, Whiteoak 1994). Car II is located to
the north of $\eta$ Car and Car I is located towards the western dark
lane, just west of Tr 14. The dynamics of the ionised gas in this region
have been studied via hydrogen recombination line emission (Gardner et al.
1970, Huchtmeier \& Day 1975) and H$\alpha$ and [\NII] emission
observations (Deharveng \& Maucherat 1975). The results show line splitting
towards the Car II region which has been interpreted as an expanding shell
of ionised gas.

The dark lanes consist of molecular gas and dust that are associated
with the nebula (Dickel 1974). H$_{2}$CO and OH absorption
measurements identified two optical depth maxima which were located
towards these lanes (Gardner, Dickel \& Whiteoak 1970, Dickel \& Wall
1974). Extended far-IR emission is confined there also (Harvey
Hoffmann \& Campbell 1979, Ghosh et al. 1988). There are two main CO
emission regions towards the nebula; a northern and southern cloud (de
Graauw et al. 1981, Whiteoak \& Otrupcek 1984). Both regions are part
of a much larger GMC which has a projected length of 130 pc and a mass
in excess of 5$\times$10$^{5}$ \Msun (Grabelsky et al. 1988). The area
between the southern and northern CO clouds is centred on the Keyhole
Nebula, a dense dark cloud northwest of $\eta$ Car. Here the molecular
gas exists in dense clumps of typical mass 10 \Msun\ that are separated
both in space and velocity (Cox \& Bronfman 1995).

The picture used to describe the Carina complex is one in which the massive
star clusters, Tr 14 and Tr 16, are interacting strongly with the molecular
cloud from which they formed.  It is generally accepted that the photons
from Tr 14 and Tr 16 are responsible for the ionised emission of Car I and
Car II respectively, and that their strong stellar winds are producing the
general expansion of the nebula. 

We are undertaking a multi-wavelength study of the Carina Complex in order
to study the detailed kinematics and distribution of the molecular and
ionised gas and to look for further evidence of ongoing star
formation. Here we present the results of initial observations of the \CO\
emission. CO emission is thermalised in both low- and high-density gas and
therefore is suitable for tracing the overall distribution and velocity
structure of the molecular cloud. It also can pinpoint any `CO
hot-spots'. These are warm molecular cores where stars could possibly be
forming.
 
\section{Observations}

Observations of the {\CO} transition at 115.271 GHz were made during
two periods, 1996 April-October and 1997 April-June, using the Mopra
Antenna, operated by the Australia Telescope National Facility,
CSIRO. At this frequency the effective diameter of the antenna is 15
m, producing a half-intensity beam width of 43 arcsec. A cryogenic
3-mm SIS receiving system was used, for which a receiver
temperature (SSB) of 130 K was assumed for all observations.

Initially observations were made on a 2 arcmin grid covering an area
of 1.5$\times$2 deg$^{2}$. This produced a coarsely sampled map which
allowed us to trace the overall molecular-cloud distribution. A total
of 7000 pointings were obtained. Follow-up observations were then made
on a finer but still under-sampled grid of 1 arcmin towards areas of
interest. We used an observing technique with a reference position
offset by 2 degrees to avoid a contribution from off-source CO
emission. One-minute integrations were made at each `source' position,
with a five-minute reference observation every thirty minutes.

The spectra were obtained using a digital correlator which was configured
into 2 bands, each consisting of 1024 channels extending over 64 MHz. One
band was tuned to the CO transition, giving a velocity resolution of 0.20
\kms. For each observing run the band was centred on a radial velocity of
-20 \kms\ (LSR). The second band was tuned to the 86-GHz SiO transition.  A
pointing accuracy of better than 10\arcsec\ was obtained by periodically
observing SiO masers near Carina.

At a wavelength of 2.6-mm, atmospheric absorption is significant. We
adopted the `chopper-wheel' method (Ulich \& Haas 1976) to correct for
this attenuation. By measuring the signal from a black-body absorbing
paddle every 30 mins by means of a method described by Hunt (1997),
observed intensities were converted to a temperature scale corrected
for atmospheric effects. To provide a temperature scale compatible
with that used at the 15-m Swedish ESO Sub-millimetre Telescope
(SEST), we observed the SEST calibration sources in Orion and M17 on a
daily basis throughout each observing period and compared the CO
temperatures to those obtained with the SEST. To make the temperature
scales consistent an average scaling factor of 1.5 was applied to 
the Mopra data. The final intensity calibration is accurate to 10-15
percent.

The data were processed using a combination of data reduction packages on
ATNF facilities. For the preliminary reduction simple baselines were
subtracted from the individual spectra and Hanning smoothing was
applied. The frequency scale was corrected to velocity (LSR) without a
diurnal variation correction, introducing a maximum error of $\pm$ 0.5
\kms. A spectral cube (RA, Dec, velocity) was made using a Gaussian
algorithm to interpolate between each spatial point. A smoothing radius of
0.8 arcmin was applied, giving an effective FWHM beam of 1.75 arcmin.

\section{Results \& Discussion}

The results from our initial observations mapped on a 2 arcmin grid
show the CO emission to extend over 3 degrees in the
northwest-southeast direction. This is in agreement with the Carina
GMC identified in the Columbia survey (Grabelsky et al. 1981). Our
higher resolution map reveals a very clumpy morphology with two strong
emission areas; one centred on Car II and the other towards a region
northwest of the optical emission. Here we present the results from
the better sampled set of observations on a 1 arcmin grid towards the
Car II region.

A grey-scale and contour image of the \CO\ emission integrated over a
velocity range of -35 to 0 \kms\ is shown in Fig 1. The grey-scale
boundaries show the region that was observed. Clearly two main
emission regions are identified; a large region of emission covering
the north-western part of the map and a region of fainter emission
located in the south-east corner. These are the northern and southern
emission clouds first distinguished by de Graauw et al. (1981) from
observations of $^{12}$CO(2--1) emission using a beamwidth of 2.2
arcmin and sampled on a 4 arcmin grid.

The northern cloud contains the strongest emission. \CO\ observations
of this region were first made by Whiteoak \& Otrupcek (1984) using a
2.8\arcmin\ beam revealing three emission concentrations. The higher
resolution data presented here clearly show a more complicated
morphology. Most of the strong emission is located in a large central
region with branches of fainter emission extending outwards from
it. One emission branch extends westwards and then expands outwards
past the observed region. This emission is part of the larger GMC
which extends over 2 degrees to the northwest. The strongest
integrated emission is located in a small concentration at the
southern part of the central emission region. Here the integrated
intensity reaches a maximum of 2.0$\times$10$^{5}$ K/Beam$\times$m
s$^{-1}$. The southern cloud is a region of fainter emission, breaking
up into small clumps at its southern edge.

The northern and southern cloud are separated by a region where the
integrated CO emission is below 10 percent of the peak integrated
intensity. This region is centred on the keyhole and Car II region.  We did
detect CO emission from the small-scale clumps that were found by Cox \&
Bronfman (1995); however the integrated intensity map shown in Fig 1 is not
sensitive to such small-scale distributions.

Our results have identified many small CO concentrations in both the northern
and southern cloud. These could be dense, warm clumps where star formation
may next occur. In fact, Megeath et al. (1996) have detected very faint
reddened stars towards one of the CO clumps in the southern cloud. If these
are very recently formed stars as suggested then this is the first
evidence of ongoing star formation within the Carina nebula. We are
currently carrying out further observation towards the many clumps defined
by this initial set of observations to look for further evidence of ongoing
star formation.

Some sample spectra taken towards nine positions throughout the nebula are
shown in Fig 2. The first three spectra (a-c) are representative of the
southern cloud region. The emission is largely at a velocity of -27 \kms\
with a second component at -20 \kms\ seen at the western edge of the
cloud. The remaining six spectra (d-i) are taken at positions in the
northern cloud. Most of the emission in this northern region is found
between velocities of -27 and -8 \kms. The southern and eastern part of the
cloud (d,e) has emission largely at -27, -18 and -10 \kms\ whereas for the
northern and western region the main emission components are at -20 and -10
\kms. The velocities presented here agree with those obtained by de Graauw
et al. (1981).

Figure 3 shows an expanded view of the area centred on the Keyhole
Nebula. The contours represent the peak integrated \CO\ emission
distribution and are superimposed on an R-band image showing
the H$\alpha$ distribution, obtained from the Digitised Sky
Survey (DSS). The figure provides a comparison between the locations of Tr
14, Tr 16 and the optical nebulosity with the distribution of
molecular gas. The keyhole region and the associated Car II radio
continuum source appear to have little interaction with the main cloud
emission. This supports the idea of the surrounding GMC being
dispersed by the stellar winds and ionising fluxes of the massive
stars in this region (Cox \& Bronfman 1995). The southern cloud
emission is confined to the eastern dust lane and coincides with an
optical depth maximum. In the north the CO emission extends over the
optical emission but the peak emission is located to the west of Tr 14
towards the western dust lane. It is coincident with a far-infrared
emission peak. Adjacent to the eastern side of the CO emission peak is
the radio continuum peak Car I. The sequence of the distinct emission
sources; ionising cluster - radio continuum peak - far-IR-peak has
been explained using a strong gradient in gas and dust density
increasing from Tr 14 to the dust lane to the west (Harvey et
al. 1979). de Graauw et al. (1981) described this region as a
blister-type region, whereby the edge of the molecular cloud is being
externally ionised by Tr 14. Our observations, which show a strong CO
concentration adjacent to the radio continuum source Car I, support
this blister-type model and clearly define the interface of the
ionised and molecular gas.

\section{Conclusion}

We have presented data from the first stage of an extensive study of the GMC
associated with the Carina nebula. The data consists of observations of
\CO\ emission which have been used to trace the overall GMC as well as
pinpoint and CO `hot-spots' or dense regions where stars could possible
form. The observations are at a higher resolution than previous studies and
reveal the clumpy nature of the northern and southern cloud regions. They
also show the positional coincidence between the far-infrared emission and
the strong CO emission towards the Car I region. This supports a
blister-type model for this region. Further observation of different
transitions are being made to better constrain the temperature and
density of the molecular gas in this interesting region.

\section{Acknowledgements}

We would like to thank the ATNF Receiver Group for tuning the system before
remote-tuning was available and for their technical support during all the
observations. This project has been funded by research grants from
the `Small Grants' ARC scheme. KJB acknowledges the support of an APA
award.

\section*{References}

\reference Cox P., Bronfman L., A\&A, 299, 583
\reference Deharveng L., Maucherat M., 1975, A\&A, 41, 27
\reference Dickel H. R., 1974, A\&A, 31, 11
\reference Dickel H. R., Wall J. V., 1974, A\&A, 31, 5
\reference DSS; for information see http:$//$skyview.gsfc.nasa.gov$/$cgi-bin$/$surv$\_$comp.pl$?$dss
\reference Feinstein A., 1995, RevMexAA, 2, 57
\reference Gardner F. F., Morimoto M., 1968, ApJ, 21, 881
\reference Gardner F. F., Milne D. K., Mezger P. G., Wilson T. L., 1970,
A\&A, 7, 349
\reference Gardner F. F., Dickel H. R., Whiteoak J. B., 1973, A\&A, 23, 51
\reference Ghosh S. K., et al 1988, ApJ, 330, 928
\reference de Graauw T., et al 1981, A\&A, 102, 257
\reference Grabelsky D. A., Cohen R. S., Bronfman L., Thaddeus P., 1988, ApJ, 331, 181
\reference Harvey P. M., Hoffmann W. F. Campbell M. F., 1979, ApJ, 227, 114
\reference Huchtmeier W. K., Day G.A., 1975, A\&A, 41, 153
\reference Hunt M., 1997, submitted to PASA
\reference Megeath S. T., Cox P., Bronfman L., Roelfsema P. R., 1996, A\&A, 305, 296
\reference Retallack D. S., 1983, MNRAS, 204,669
\reference Tovmassian H. M., 1995, RevMexAA, 2, 83
\reference Ulich B. L., Haas R. W., 1976, ApJ, 30, 247
\reference Walborn N. R., 1995, RevMexAA, 2, 51
\reference Whiteoak J. B., Otrupcek R. E., 1984, PASA, 5(4), 552
\reference Whiteoak J. B. Z., 1994, ApJ, 429, 225

\newpage
\begin{figure}
\begin{center}

\caption{The grey-scale and contour image of the \CO\ emission
integrated over a velocity range of -30 to 0 \kms. The contour levels
are 16, 22, 28, 34, 40, 46, 52, 58, 64, 67, 70, 82, 88, 94, 100
percent of the peak intensity, 2.0 $\times$10$^{5}$
K$/$Beam$\times$m s$^{-1}$. The faint grey emission shows the total
observing area.}
\label{fig1}
\vspace{1.0cm}
\psfig{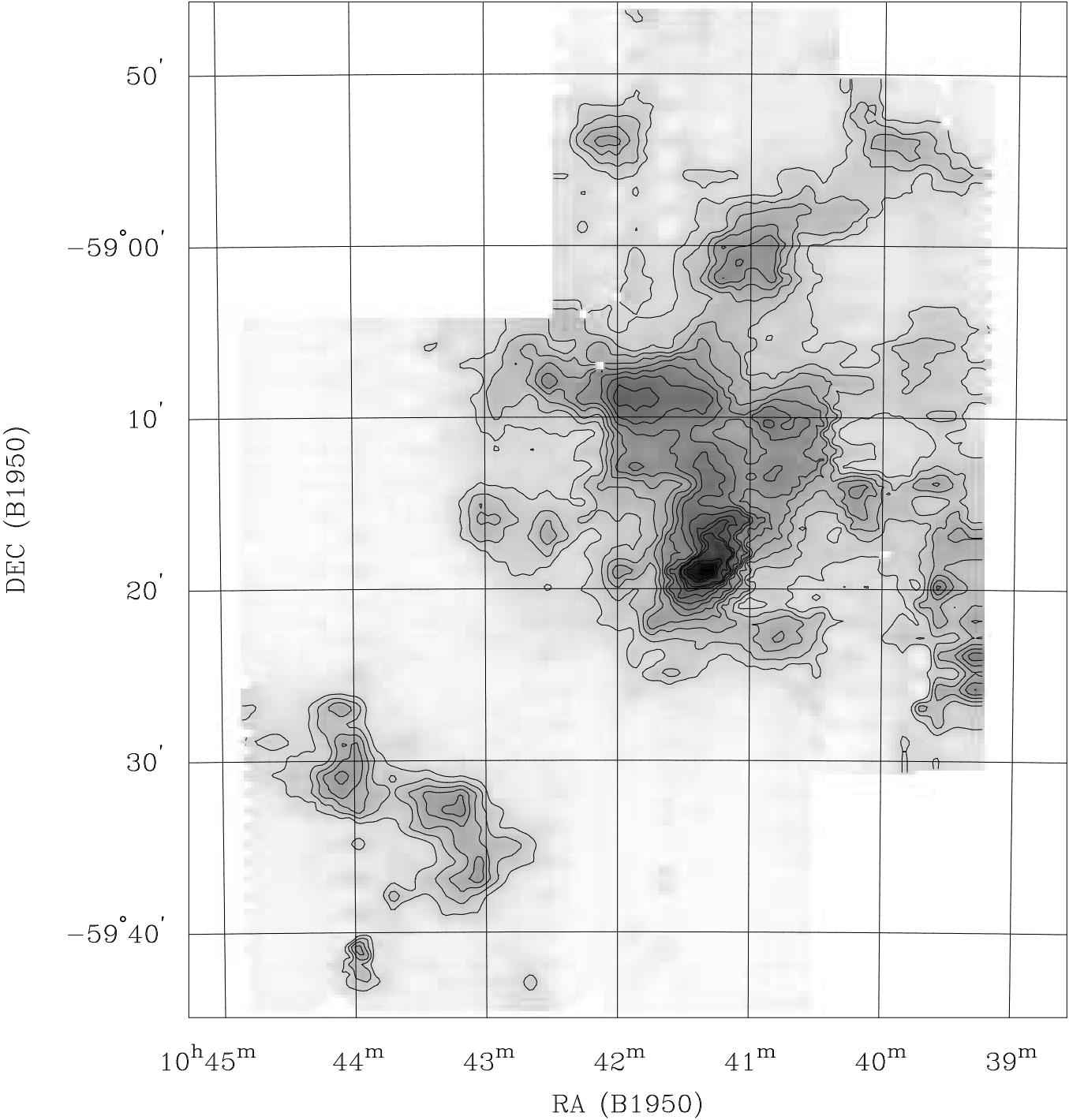}
\end{center}
\end{figure}

\newpage
\begin{figure}
\begin{center}
\caption{Typical spectra of the \CO\ emission at nine positions throughout
the northern and southern cloud. Positions are given in B1950 coordinates.}
\vspace{-2.0cm}
\psfig{file=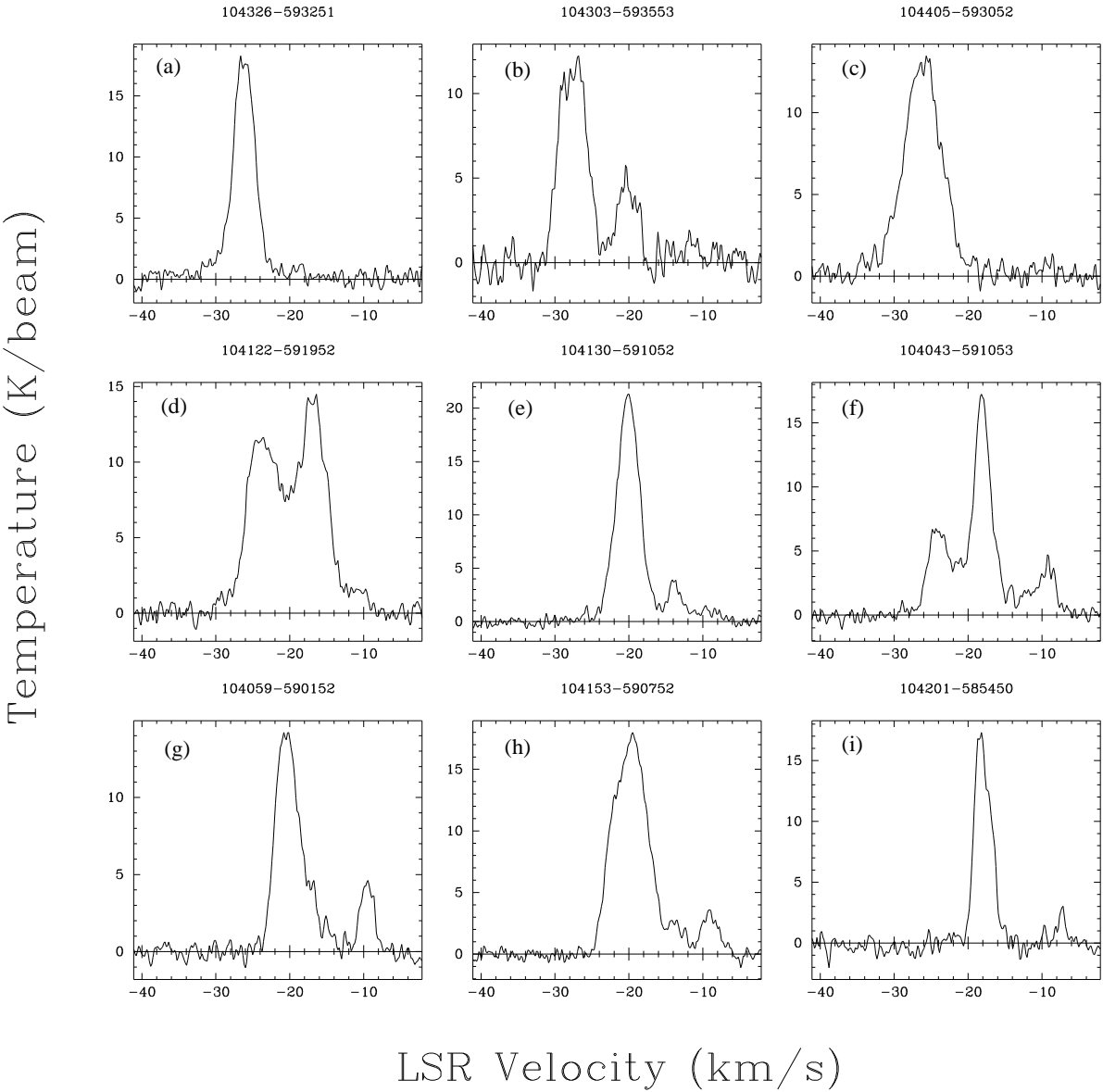,width=18cm}
\label{fig2}
\end{center}
\end{figure}

\newpage
\begin{figure}
\begin{center}
\caption{A contour representation of the \CO\ emission integrated
over a velocity range of -30 to 0 \kms, superimposed on an R-band image. The contour levels are 10, 16, 22, 28, 34, 40, 46, 52, 58,
64, 67, 70, 82, 88, 94, 100 percent of the peak intensity, 2.0
$\times$10$^{5}$ K$/$Beam$\times$m s$^{-1}$. Tr 14 is concentrated in the
northwest corner of the map and Tr 16 is situated near the centre.}
\vspace{1.0cm}
\psfig{file=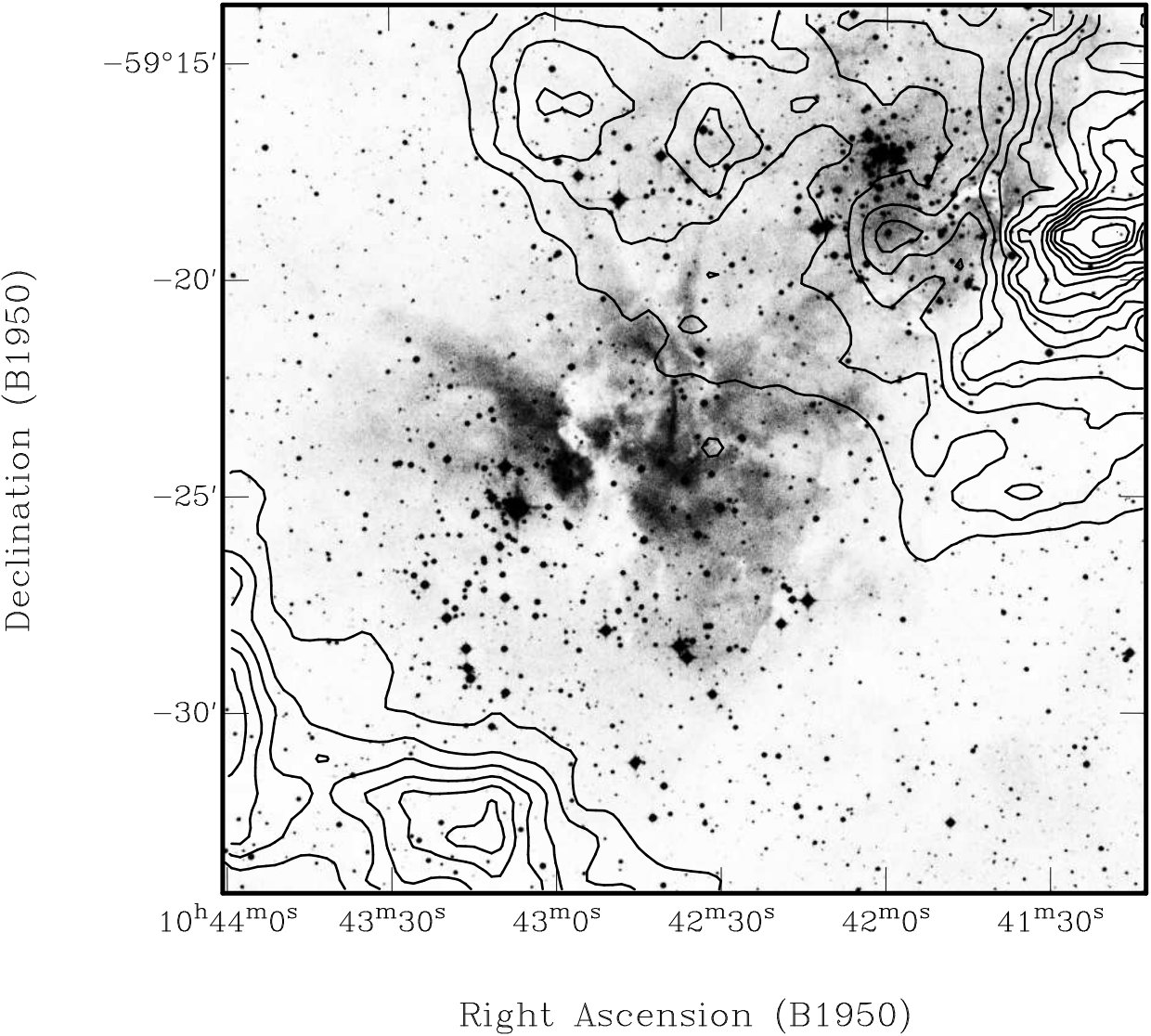,width=18cm}
\label{fig3}
\end{center}
\end{figure}

\end{document}